\documentclass[a4paper,twocolumn,11pt]{quantumarticle}
\pdfoutput=1
\usepackage[utf8]{inputenc}
\usepackage[english]{babel}
\usepackage[T1]{fontenc}
\usepackage{amsmath}
\usepackage{hyperref}

\usepackage{tikz}
\usepackage{lipsum}
\usepackage{cite}
\usepackage{array}

\begin{document}

\title{Long-time Bell states of waveguide-mediated qubits via continuous measurement}

\author{Huiping Zhan}
\affiliation{Department of Physics, Huazhong Normal University, Wuhan 430079, China}
%\orcid{0000-0002-2445-2701}
\author{Huatang Tan}
\email{tht@mail.ccnu.edu.cn}
\affiliation{Department of Physics, Huazhong Normal University, Wuhan 430079, China}

\maketitle
\begin{abstract}
  The generation of Bell states of distant objects is of importance for constructing quantum networks. Previous studies have revealed that transient or intermittent Bell states can be generated between remote qubits by exploiting time-continuous measurement on the environments of the systems, e.g., photon counting or homodyne detection. In this paper, we consider a new scheme for achieving long-time sustainable Bell states of two distant qubits mediated by a one-dimension waveguide via continuous photon counting and homodyne detection. In both of cases, different Bell states can be present for different initial states in the long-time regime. Specially, in the former case, we find that a cyclic jump among Bell states can be formed once the first photon is registered, and more interestingly in the latter case, any steady Bell state can be achieved independent of detection efficiency.
\end{abstract}

\section{Introduction}
Apart from fundamental interests of research, entanglement has nowadays become a core resource of quantum informatics \cite{man}. Bell states, as maximally entangled two-qubit states, have perfect quantum correlations and are therefore especially important for realizing various high-efficient quantum tasks, such as quantum teleportation \cite{cha1}. Many kinds of protocols have been designed to generate Bell states in different quantum matters, like atomic systems, quantum dots, superconducting qubits, and magnon-photon system \cite{bel1, bel2, bel3, bel4}. Compared to such short distance entanglement, long distance entanglement is of importance for distributing and transmitting quantum information among distant quantum nodes in quantum networks \cite{net1,net2}. To this end, new light has recently been shed on waveguide QED (quantum electromagnetic dynamics) systems, which are excellent integration platforms for generating long-distance entanglement and building waveguide quantum networks, owing to their characteristics of controllable interactions between matters and light and combining them with open propagation directions \cite{to1,to2,too1,too2,too3,too4,too5,too6,too7,to3,to4,to5,to6,to7,to8}.

The studies for generating entanglement between two distant quantum emitters (qubits) mediated by e.g. photonic, plasmonic, and magnonic waveguides have been
carried out \cite{phw1,phw2,plw1,plw2,mw1}. However, mixed entangled states are merely resulted, due to unavoidable decoherence process, such as spontaneous emission. As we know, spontaneous emission process can be envisaged as an ensemble of trajectories of time-continuous quantum weak measurements on the environment induced the docohering process \cite{Hcar,cur1,Jz,SV,Car}, pure Bell states of some trajectory may be generated via measurements, e.g., continuous photon counting and homodyne detection \cite{me1,me2,me3,phi1,phi2,phi3,ond}. For example, it has recently been shown that short-time Bell states of quantum trajectories can be achieved by continuously homedyning the outputs of a beam splitter on which the spontaneous fluorescences from two qubits are incident \cite{phi1}. It was also shown that entangled states of two remote qubits connected with fiber can be achieved via homodyne detection \cite{ond}. The measurement provides information on the total spin of the two qubits such that the entanglement can be postselected. Experimentally, the entanglement of quantum trajectories of homodyne detection on two distant qubits has been demonstrated \cite{are}. Nevertheless, the entangled states just appear in the transient regime and moreover the Bell states merely exist at some time points, since the qubits initially prepared in excited states inevitably relax to ground states. To pull the qubits back to the excited states, one can employ classical strong driving field. For instance, Zhang $et$ $al.$ \cite{xhh} recently have proposed a scheme for the heralded generation of Bell state of waveguide-mediated qubits driven by classical laser via continuous photon counting. However, the Bell state is merely present in a intermittent manner, since it takes time for the qubits back to excites states, and moreover the conditional Bell state almost collapses even when the detection efficiency deviates from unit slightly.

In this paper, we consider a scheme for achieving long-term sustainable Bell states of two distant waveguide-mediated qubits via photon counting or homodyne detection. The system under our consideration consists of two laser-driven identical emitters in the $\Lambda$ configuration which are coupled to a one-dimension waveguide via off-resonant Raman scattering. The outputs of the waveguide are subject to continuous photon counting or homodyne detection. In both of cases, it is shown that different types of Bell states can be realized in the long-time regime. For the photon-counting case, we find that a cyclic jump among Bell states is formed once the first photon is detected, meaning that different Bell states appear alternately, conditioned on the occurrence of subsequent photon-detection events.  While for the homodyne detection case, we show that initial-state-dependent Bell states can be obtained in the regime of steady states, with a probability of fifty percent. Moreover, this is independent of homodyne detection efficiency.

The remainder of the paper is organized as follows. In Sections II and III,  the system and the working equations are presented. In Section VI, we investigate in detail the properties of the entanglement of trajectories via photon counting and homodyne detection. In the last section, the summary is given.

\begin{figure}[t]
\centering
\includegraphics[bb=210 160 400 560,scale=0.52]{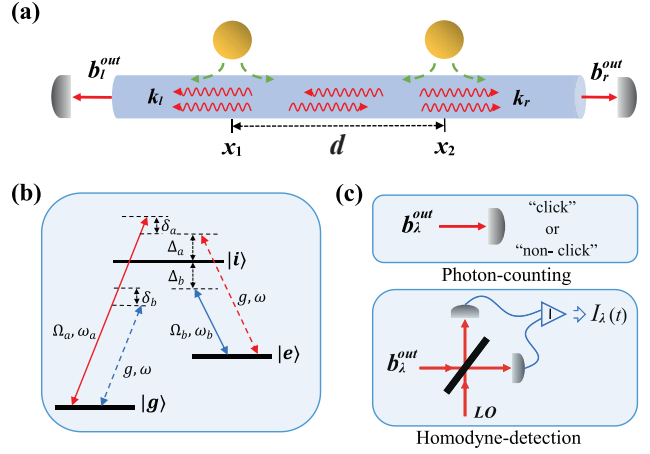}
\caption{(a) A schematic plot of the system. Two identical $\Lambda$-type emitters are located at the positions $x_{1,2}$ along the waveguide and coupled to the left and right propagating waveguide modes $b_{l,r}$ with wave numbers $k_{l,r}$. The output fields $b_{l,r}^{out}$ of the waveguide are subject to photon counting or homodyne detection, as shown in (c). (b) Energy level diagram of the $\Lambda$-type emitters, in which the ground state $|g\rangle$ and excited state $|e\rangle$ are coupled to the auxiliary state $|i\rangle$ through the driving fields of frequency $\omega_{a,b}$ and the waveguide modes of frequency $\omega$.}
\label{1}
\end{figure}
\section{system and equations}
%\subsection{Model and Hamiltonian }
As schematically shown in Fig. \ref{1}~(a), our system consists of two identical emitters in the $\Lambda$ configuration and coupled to a one-dimensional waveguide with distance $d$. Each emitter is driven by classical fields with frequencies $\omega_i~(i=a,b)$ and simultaneously coupled to the left and right transmitting  waveguide modes $\hat b_\lambda~(\lambda=l,r)$ with frequency $\omega$ and wave vector $k_\lambda$, as depicted in Fig. \ref{1}~(b). In the interaction picture with respect to the free Hamiltonian of the system, the Hamiltonian of the whole system can be written as
\begin{align}
	\hat H_{I}(t)&=i\sum_{\lambda=l,r}\sum_{j=1,2}\int \Big[g\hat b_{\lambda}(\omega)\hat \sigma_{ig}^je^{ik_\lambda x_j}e^{i(\Delta_b+\delta)t}\nonumber\\
	&+g\hat b_{\lambda}(\omega)\hat \sigma_{ie}^je^{ik_\lambda x_j}e^{-i\Delta_at}+\Omega_a\hat \sigma_{ig}^j e^{-i(\Delta_a+\delta)t}\nonumber\\
	&+\Omega_b\hat\sigma_{ie}^j e^{i\Delta_bt}\Big]d\omega+h.c.,
	\label{hba}
\end{align}
describing the coupling of the emitters at positions $x_{1,2}$ to the driving fields and the waveguide (as a reservoir).  Here, the detunings $\Delta_a=\omega-\omega_{ie}$, $\Delta_b=\omega_{ie}-\omega_b$, $\delta_a=\omega_a-\omega_{eg}-\omega$, and $\delta_b=\omega_b+\omega_{eg}-\omega$. Consider the situations: the detuning $|\Delta_i|\gg|\Omega_i|$, such that the excited state $|i\rangle$ can be adiabatically eliminated; the dispersive atom-waveguide interaction from the adiabatical elimination can be ignored for the condition $|\Omega_i|\gg|g|$. The effective Hamiltonian  $\hat{\widetilde{H}}=\hat{\widetilde{H}}_0+\hat{\widetilde{H}}_I$ is derived as
\begin{subequations}
	\begin{align}
		\hat{\widetilde{H}}_0&=\sum_{j=1,2}\frac{1}{2}\widetilde{\omega}\hat\sigma_z^{j},\\
		\hat{\widetilde{H}}_I&=i\sqrt{\frac{\gamma}{2\pi}}\sum_{\lambda=l,r}\sum_{j=1,2}\int  \hat b_\lambda^\dag(\omega)(\hat \sigma_j+\hat \sigma_j^\dag)\nonumber\\
&~~~~\times e^{-i(k_\lambda x_j+\delta t)}d\omega+h.c.,
		\label{hef}
	\end{align}
\end{subequations}
on the conditions $\delta_a=-\delta_b=\delta$ and $|\Delta_a/\Delta_b|=|\Omega_a/\Omega_b|$ for which $\omega$ is replaced by $\omega_c$ approximately since we are interested in a narrow bandwidth centered at $\omega_c$. Here, $\hat{\widetilde{H}_0}$ is the free Hamiltonian of a two-level emitter with the transition frequency $\widetilde\omega=-\sum_i\frac{4|\Omega_i|^2}{\Delta_i}$ between the energy levels $|e\rangle$ and $|g\rangle$ (as our qubit). $\hat{\widetilde{H}}_I$ describes effective interaction between the qubits and the waveguide, with $\gamma/2\pi=4\mid g\Omega_i/\Delta_i\mid^2$ and the lowering operator
$\hat \sigma_j=\hat \sigma_{ge}^j(\hat\sigma_j^\dag=\hat\sigma_{eg}^j)$.  For initial vacuum of the waveguide modes, the master equation for the density operator $\hat\rho_{a}$ of the qubits under the Born-Markovian approximation can be derived as
%\begin{align}
%\frac{d}{dt}\hat\rho_{a}=-i[\hat H_m,\hat\rho_{a}]+\gamma\sum_{\lambda=l,r}\mathcal D[\hat J_\lambda]\hat\rho_{a}
%\frac{d}{dt}\hat\rho_{a}=+\gamma\sum_{\lambda=l,r}\mathcal D[\hat J_{\lambda\pm}]\hat\rho_{a}
%\label{drhoac}
%\end{align}
%where $\hat H_m=2\gamma\sin (kd)(\hat J_1^\dag\hat J_2+\hat J_2^\dag\hat J_1)$, with $\hat J_{j=1,2}=(\hat \sigma_{j}+\hat \sigma_{j}^\dag)/\sqrt{2}$. Here $\mathcal D[\hat O]\hat\rho= \hat O\hat \rho  \hat O^\dag-\{ \hat O^\dag  \hat O,\hat\rho\}/2$.
%The second terms in Eq. (\ref{drhoac}) characterize the collective dissipation of $\hat J_{\lambda}=\sum_{j=1,2}e^{ik_\lambda x_j}\hat J_j$ of the emitters ($k_\lambda$ is the wave vector and $k_r=-k_l=-k$) into the left and right propagating continua of the waveguide at the effective decay rate $\gamma$.
%In the case of  $kd=2n\pi$ or $(2n+1)\pi$ for integer number $n$, Eq. (\ref{drhoac}) can be reduced as
\begin{align}
	%\frac{d}{dt}\hat\rho_{a}=-i[\hat H_m,\hat\rho_{a}]+\gamma\sum_{\lambda=l,r}\mathcal D[\hat J_\lambda]\hat\rho_{a}
	\frac{d}{dt}\hat\rho_{a}=-i\big[\hat{\widetilde{H}}_0,\hat\rho_{a}\big]+\gamma\sum_{\lambda=l,r}\mathcal D\big[\hat J_{\lambda\pm}\big]\hat\rho_{a},
	\label{rho}
\end{align}
with the distance $kd=2n\pi$ or $(2n+1)\pi$ for integer number $n$ and $k_r=-k_l=k$. The symbol $\mathcal D[\hat O]\hat\rho= \hat O\hat \rho  \hat O^\dag-\{ \hat O^\dag  \hat O,\hat\rho\}/2$, where the operators $\hat J_{\lambda\pm}=\hat J_1\pm\hat J_2$ respectively for $kd=2n\pi$ and $kd=2(n+1)\pi$, with $\hat J_{j=1,2}=(\hat \sigma_{j}+\hat \sigma_{j}^\dag)/\sqrt{2}$.
Eq. (\ref{rho}) effectively describes the dissipative-driven collective dynamics of two qubits immersed in a one-dimension bosonic environment.
%which can be
%equivalently characterized by the interaction
%\begin{align}
%\hat{\widetilde H}_I=i\sqrt{\gamma}\big[\hat J_{\lambda\pm} \hat %b_\lambda^\dag(\omega)-\hat J_{\lambda\pm}^\dag \hat b_\lambda(\omega)\big].
%\label{effhm}
%\end{align}
% In this way, the input-output relations for the waveguide can be given by
%\begin{align}
%\hat b_\lambda^{out}(t)=\hat b_\lambda^{in}(t)+\sqrt{\gamma}\hat J_{\lambda\pm},
%\label{bout}
%\end{align}
%where the input vacuum noise operators satisfy $[\hat b_\lambda^{in}(t),\hat b_\lambda^{in\dag}(t')]=\delta(t-t')$. It is shown from Eq.(\ref{bout}) the information of the spin of the qubits measurement can be traced by measuring the waveguide output fields $\hat b_\lambda ^{out}(t)$.
Note that the time delays is neglected by assuming that the time scale $T_1=\gamma^{-1}$ on which the system evolves is much larger than the photon travelling time between the two emitters.
%Then, by tracing out the degrees of freedom of the waveguide, the reduced system density matrix $\rho(t)$  can be described by the master equation:

%If the total hamiltonian in the first term of Eq. (\ref{rho}) is 0, then Eq. (\ref{rho}) will have the steady-state solution $\rho_\infty=|\Psi\rangle\langle\Psi|$, in which $S_\lambda|\Psi\rangle=0$.

\section{Time-continuous measurements}
We consider continuous measurement on the waveguide's outputs
$\hat b_\lambda^{out}(t)=\hat b_\lambda^{in}(t)+\sqrt{\gamma}\hat J_{\lambda\pm}$,
where the input vacuum noise satisfy $[\hat b_\lambda^{in}(t),\hat b_\lambda^{in\dag}(t')]=\delta(t-t')$. It is shown that measurements can gain information about the spin of the qubits, which render stochastic evolution of the system's state, conditioned on the measurement records \cite{Hcar}. The master equation (\ref{rho}) can be unraveled in a completely different manner, such as photon-counting detection or homodyne detection, which lead to jumpy or diffusive quantum trajectories, respectively.

%\subsection{Photon counting}
For the case of photon counting, as shown in Fig. \ref {1}~(c), the photodetector clicks every time, indicating its registering a single
photon emitted from the left or right outputs. With a generic detection efficiency $\eta_\lambda$ ($\eta_\lambda\in [0,1]$), the stochastic master equation for the conditional density matrix $\hat \rho_p$ is given by \cite{cur1}
\begin{align}
	d\hat \rho_p&=-i[\hat{\widetilde H}_0,\hat \rho_p]dt-\sum_{\lambda=l,r} \gamma\mathcal{ H}[\frac{\eta_\lambda}{2}\hat J_{\lambda\pm}^\dag \hat J_{\lambda\pm}]\hat \rho_pdt\nonumber\\
	&~~~+\sum_{\lambda=l,r} \gamma\mathcal { D}[\sqrt{(1-\eta_\lambda)}\hat J_{\lambda\pm}]\hat \rho_p dt\nonumber\\
	&~~~+\sum_{\lambda=l,r} \mathcal { G}[\sqrt{\eta_\lambda}\hat J_{\lambda\pm}]\hat \rho_pdN_\lambda,
	\label{drhop}
\end{align}
%for jump trajectories of state $\rho_p$,
with the symbols $\mathcal{H}[\hat O]\hat \rho=\hat O\hat \rho+\hat \rho \hat O^\dag-\mathrm{Tr}[\hat O\hat \rho+\hat \rho \hat O^\dag]\hat \rho$ and
$\mathcal{G}[\hat O]\hat \rho=\frac{\hat O\hat \rho \hat O^\dag}{\mathrm{Tr}(\hat O\hat \rho \hat O^\dag)}-\hat \rho$. The stochastic variable $dN_\lambda(t)$ denotes the measurement results ($dN_\lambda(t)=0$ or $dN_\lambda(t)=1$) during a infinitesimal time interval $dt$. For perfect detection $\eta_\lambda=1$, when a photon is registered ($dN_\lambda(t)=1$), the system's state jumps to
$|\psi_1(t+dt)\rangle\rightarrow\sum_{\lambda=l,r} \sqrt{\gamma}\hat J_{\lambda\pm} |\psi(t)\rangle$
from the state $|\psi(t)\rangle$ at the time $t$, with the probability $\langle dN_\lambda(t)\rangle=\langle \psi(t)|\hat J_{\lambda\pm}^\dag \hat J_{\lambda\pm}|\psi(t)\rangle dt $. When no photon is registered ($dN_\lambda(t)=0$), the system's state collapses into $
|\psi_0(t+dt)\rangle\rightarrow \big\{1-\big[\sum_{\lambda=l,r}\frac{\gamma }{2}\hat J_{\lambda\pm}^\dag \hat J_{\lambda\pm}+i\hat{\widetilde H}_0\big]dt\big\} |\psi(t)\rangle$.

For the case of continuous homodyne detection on the waveguide's outputs, the stochastic master equation for the density operator $\hat \rho_{c}$ is \cite{cur1}
\begin{align}
	d\hat \rho_{c}&=-i[\hat{\widetilde H}_0,\hat\rho_c]dt+\sum_{\lambda=l,r}\gamma\mathcal D[\hat J_{\lambda\pm}]\hat\rho_cdt\nonumber\\
	&~~~~+\sum_{\lambda=l,r}\sqrt{\frac{\eta_\lambda\gamma}{2}}\mathcal{H}[\hat J_{\lambda\pm}]\hat\rho_cdW_\lambda(t),
	\label{drhoc}
\end{align}
conditioned on the detection currents
\begin{align}
	I_{\lambda }(t)dt&=\sqrt{\eta_\lambda\gamma}\langle \hat J_{\lambda\pm}+\hat J_{\lambda\pm}^\dag\rangle dt +dW_{\lambda }(t),
	%I_{\lambda y}(t)&=\frac{i}{\sqrt{2}}\langle S_\lambda-S_\lambda^\dag\rangle +\xi_{\lambda y}(t),
\end{align}
\label{mbs}
where $\eta_\lambda$ are homodyne detection efficiencies and $dW_\lambda(t)$ the standard Wiener increments with mean zero and variance $dt$.

\section{results and discussion }
In this section,we investigate in detail the properties of the entanglement between the two qubits  via photon-counting and homodyne detection. The stochastic master equations (\ref{drhop}) and (\ref{drhoc}) are numerically solved with using the \textsc{python} package QuTiP\cite{JR1,JR2}. The degree of entanglement between the two emitters is measured by the concurrence \cite{curr}
\begin{align}
	\mathcal{C}(\hat \rho)=\mathrm {max}\big\{0,\lambda_1-\lambda_2-\lambda_3-\lambda_4\big\},
\end{align}
for the density operator $\hat \rho$ in the basis $\{|0\rangle\equiv|g_1g_2\rangle, |1\rangle\equiv|g_1e_2\rangle, |2\rangle\equiv|e_1g_2\rangle, |3\rangle\equiv|e_1e_2\rangle\}$, where $\lambda_i$  are the  square roots of the eigenvalues, in decreasing order, of the non-Hermitian matrix $\hat\rho(\hat \sigma_y\otimes\hat\sigma_y)\hat \rho^*(\hat \sigma_y\otimes\hat\sigma_y)$.
%In addition, in our consideration the distance of two emitters $kd=2n\pi$, and the %levels in each qubits are degenerate, i.e., $\hat{\widetilde H}_0=0$.

\subsection{Bell states via photon counting}
\begin{figure}[t]
	\centering
	\includegraphics[bb=-20 -50 1000 545,scale=0.52]{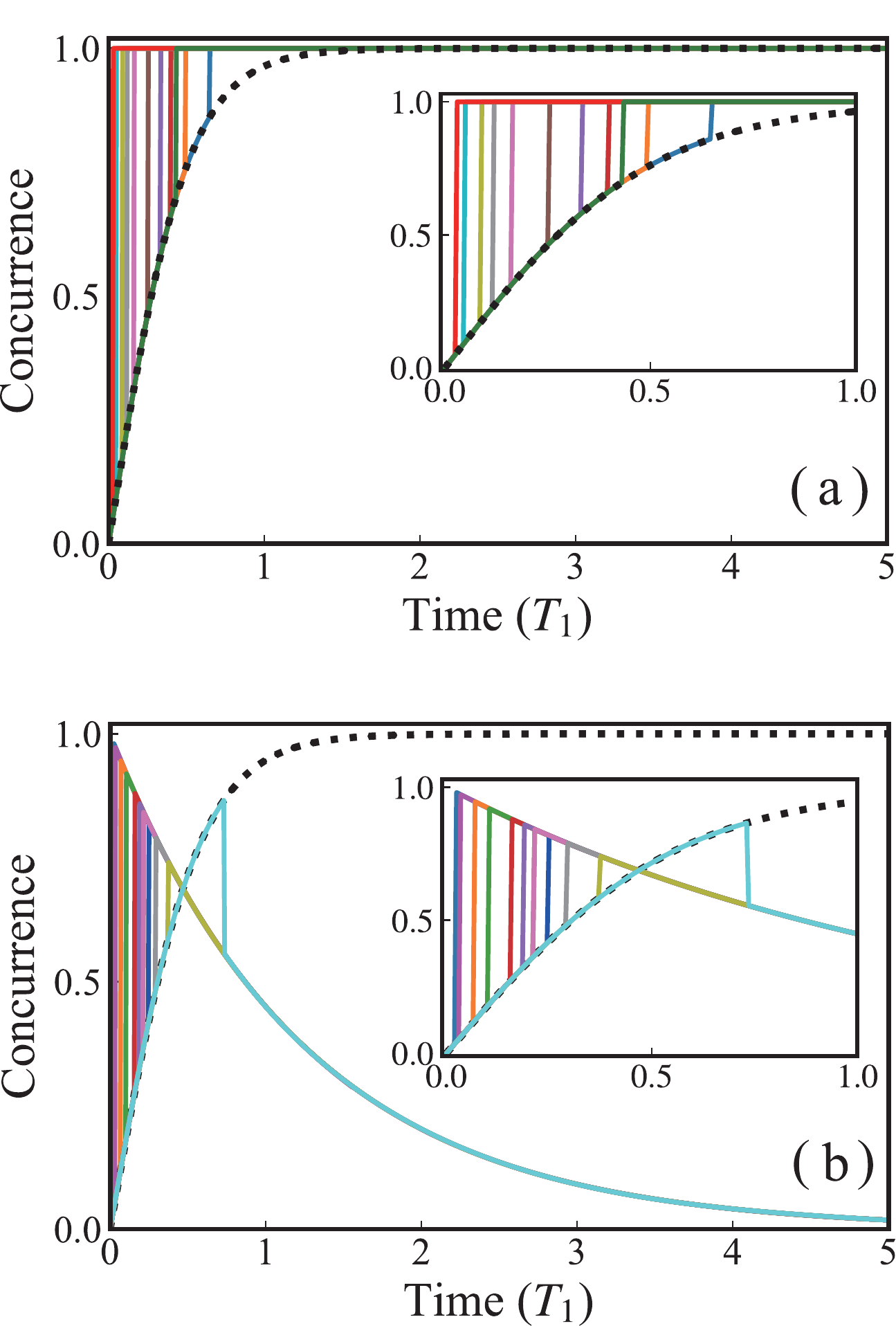}
	\caption{The concurrence of a dozen individual jump trajectories (color solid lines) initialized from $|g_1g_2\rangle$ for (a) ideal photon detection ($\eta_l=\eta_r=1$) and (b) inefficient photon detection ($\eta_l=\eta_r=0.9$). The black dotted lines represent the situation where no photon is recorded during the measurement process. The parameters $kd=2n\pi$, $T_1=\gamma^{-1}$,  and $dt=T_1/200$. }
	\label{3}
\end{figure}

\begin{figure}[t]
	\centering
	\includegraphics[bb=20 90 1000 640,scale=0.41]{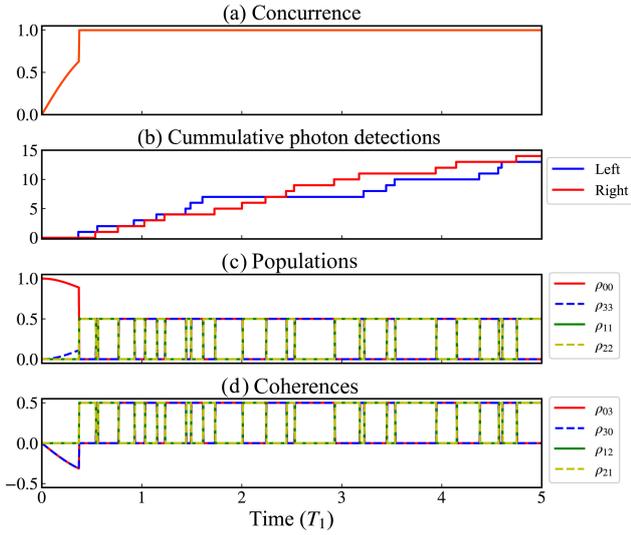}
	\caption{(a) A single jump trajectory of concurrence, (b) Cumulative photon detections at left (blue stepped-line) and right (red stepped-line) output ports, (c) populations and (d) coherences, for perfect photon detection $\eta_l=\eta_r=1$. The other parameters are the same as in Fig. {\ref 3}.}
	\label{4}
\end{figure}

In Fig.\ref{3} (a) and (b), we plot the time evolution of the concurrence of  different jump trajectories for initial ground state $|g_1g_2\rangle$ and the distance $kd=2n\pi$, with the detection efficiencies $\eta_l=\eta_r=1$ and $\eta_l=\eta_r=0.9$, respectively. The black dotted curves represent the entanglement in the case that no photons are detected during the measurement process. It is shown that before a photon is detected, the entanglement increases as time develops. This is clearly shown in Fig.\ref{4} (a) where the concurrence of a single trajectory is plotted. Furthermore, after the first photon is registered, we see that the maximal concurrence $\mathcal C=1$ can be always kept afterwards even in the long-time regime, which implies the two qubits are in pure Bell states. This can be understood as follows: when starting from the ground state $|g_1g_2\rangle$, the qubits will evolve into the entangled state of the superposition between $|g_1g_2\rangle$ and $|e_1e_2\rangle$, as a consequence of the driving from the ground state to the excited state. Specifically, before the first photon registration (conditioning the environment being in vacuum), the system is governed by the unitary operator $\hat U_0(t)=\textrm{exp}\big[(-\sum_{\lambda=l,r}\frac{\gamma}{2}\hat J_{\lambda+}^\dag\hat J_{\lambda+}\big)t\big]$, and for the initial state $|g_1g_2\rangle$ the system's state evolves into

\begin{align}\label{psi}
|\psi'(t)\rangle&=\frac{1+e^{-2\gamma t}}{\sqrt{2(1+e^{-4\gamma t})}}|g_1g_2\rangle\nonumber\\
&~~~-\frac{1-e^{-2\gamma t}}{\sqrt{2(1+e^{-4\gamma t})}}|e_1e_2\rangle,
\end{align}

with the concurrence
\begin{equation}\label{con}
\mathcal{C}(t)=\frac{1-e^{-4\gamma t}}{1+e^{-4\gamma t}}.
\end{equation}
We therefore see that the entanglement increases monotonically until the first photon is detected and the Bell state $|\Psi_-\rangle=(|g_1g_2\rangle-|e_1e_2\rangle)/\sqrt{2}$ can be achieved, conditioned no photons are detected during the time $t \gg \gamma^{-1}$.

Once a photon is registered by either detector, indicated from Eq.(\ref{psi}), the state $|\psi'(t)\rangle$ is then immediately projected onto the other Bell state  $|\Phi_+\rangle=(|g_1e_2\rangle+|e_1g_2\rangle)/\sqrt{2}$ after the first jump, since the information on which qubit emits the photon is erased. Moreover, this state will be maintained if no subsequent photon registration occurs, because the transitions $|e_i\rangle\rightarrow |g_i\rangle$ and $|g_i\rangle\rightarrow |e_i\rangle$ simultaneously take place, according to the interaction in Eq.(\ref{hef}). Until the second ``click" occurs, $|\Phi_+\rangle$ jumps to a new Bell state $|\Psi_+\rangle=(|g_1g_2\rangle+|e_1e_2\rangle)/\sqrt{2}$ and then jumps back to state $|\Phi_+\rangle$ at some time that another photon is detected again. As a result, a cyclic jump between $|\Phi_+\rangle$ and $|\Psi_+\rangle$ is formed on the condition that ``clicks" take place. The concurrence thus always keeps its maximum after a transient increases. This is also exemplified in Fig.\ref{4} (b)-(d) where the cumulative photon-detection events, the populations and coherence of the two qubits are respectively revealed for a single trajectory. It is shown that the populations of $\{\rho_{00}$, $\rho_{33}\}$ and $\{\rho_{11}$, $\rho_{22}\}$  and the corresponding coherence $\{\rho_{03}, \rho_{30}\}$ and $\{\rho_{12}, \rho_{21}\}$ take the values of $0$ and $0.5$ alternately after a transient growth from zero.

\begin{figure}
	\centering
	\includegraphics[bb=20 220 1000 520,scale=0.40]{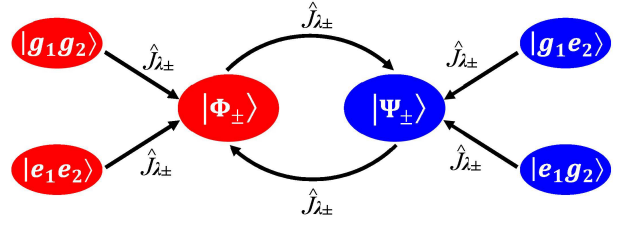}
	\caption{A diagram of cyclic jump between Bell states   $|\Phi_\pm\rangle=(|g_1e_2\rangle\pm|e_1g_2\rangle)/\sqrt{2}$ and $|\Psi_\pm\rangle=(|g_1g_2\rangle\pm|e_1e_2\rangle)/\sqrt{2}$, enabled by the jump operators $\hat J_{\lambda\pm}=\hat J_1\pm\hat J_2$ on different initial states $|g_1g_2\rangle$, $|e_1e_2\rangle$, $|g_1e_2\rangle$, or $|e_1g_2\rangle$.}
	\label{2}
\end{figure}

In fact, as illustrated in Fig.\ref{2} where the operators $\hat J_{\lambda+}=\hat J_1+ \hat J_2$ and $\hat J_{\lambda-}=\hat J_1- \hat J_2$ respectively for $kd=2n\pi$ and $kd=(2n+1)\pi$, similar state cycles between  $|\Phi_\pm\rangle=(|g_1e_2\rangle\pm|e_1g_2\rangle)/\sqrt{2}$ and $|\Psi_\pm\rangle=(|g_1g_2\rangle\pm|e_1e_2\rangle)/\sqrt{2}$ can also be formed, dependent on the initial states of $|g_1g_2\rangle$, $|e_1e_2\rangle$, $|g_1e_2\rangle$, and $|e_1g_2\rangle$. This is because of two possible channels for creating photons: one is the transition $|e_j\rangle \rightarrow |g_j\rangle$ via the interaction $\hat b_\lambda^\dag \hat \sigma_j$ and the other is $|g_j\rangle\rightarrow|e_j\rangle$ through $\hat b_\lambda^\dag\hat \sigma_j^\dag$, according to Eq.(\ref{hef}). Since the measurement is unable to distinguish from which channel the photon is created, the operator $\hat J_{\lambda\pm}$ can realize following jump processes: $\hat J_{\lambda\pm}|g_1g_2\rangle\rightarrow |e_1g_2\rangle\pm |g_1e_2\rangle$, $\hat J_{\lambda\pm}|g_1e_2\rangle\rightarrow |e_1e_2\rangle\pm |g_1g_2\rangle$,  $\hat J_{\lambda\pm}|e_1g_2\rangle\rightarrow |g_1g_2\rangle\pm |e_1e_2\rangle$, and
$\hat J_{\lambda\pm}|e_1e_2\rangle\rightarrow |g_1e_2\rangle\pm |e_1g_2\rangle$.

As discussed above, in our scheme the qubits can always be in one of maximally entangled states, due to the state cycle between $|\Phi_\pm\rangle$ and $|\Psi_\pm\rangle$. Evidently, if the interaction in Eq.(\ref{hef}) only contains the terms $\hat b_\lambda^\dag \hat \sigma_j$, as in Ref. \cite{xhh}, the first ``click" can also herald a Bell state $|\Phi_\pm\rangle$ for initial state $|g_1g_2\rangle$. However, the achieved Bell state will jump back to the ground state due to spontaneous emission. It takes time to excite the qubits back to the excited states and then a subsequent ``click" projects the qubits again into another Bell state. This is repeated for perfect detection efficiency. In the present scheme, because there exists the engineered terms $\hat b_\lambda^\dag \hat \sigma_j^\dag$, the system can always be kept in a Bell state for perfect detection.
\begin{figure*}[t]
	\centering
	\includegraphics[bb=270 10 860 720,scale=0.4]{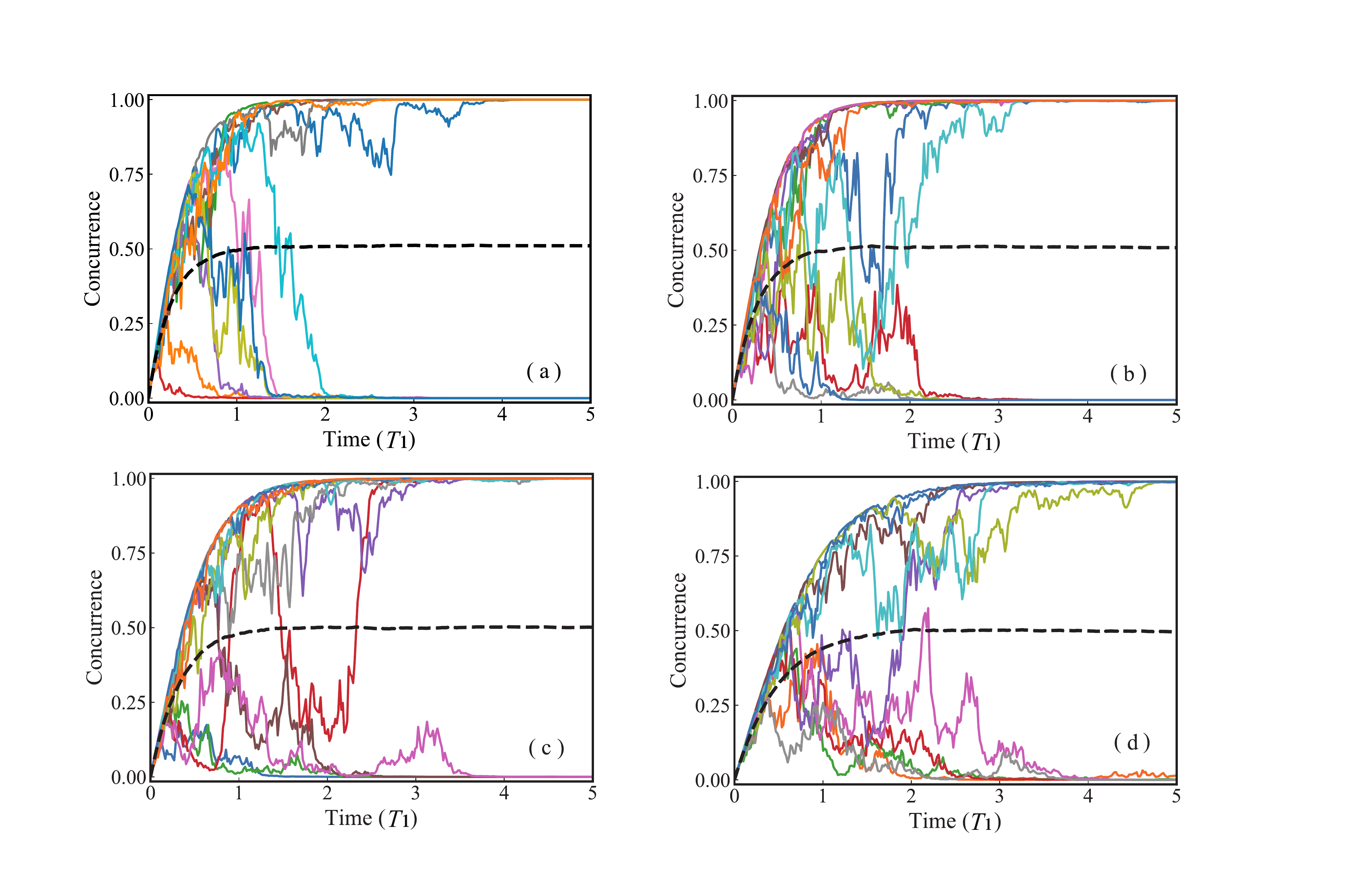}
	\caption{The evolution of the concurrence of individual trajectories (color solid lines) and the average concurrence over an ensemble of $2000$ diffusive trajectories (black dashed line) when the right outport is homodyned for initial state $|g_1g_2\rangle$, with the detection efficiency $\eta_r=1$ in (a), $\eta_r=0.9$ in (b), $\eta_r=0.75$ in (c), and $\eta_r=0.5$ in (d). The other parameters $kd=2n\pi$, and $T_1=\gamma^{-1}$.  }
	\label{5}
\end{figure*}

In Fig.\ref{3} (b), the entanglement of trajectories for finite detection efficiency is plotted. We see that under the inefficient photon detection, the concurrence of individual trajectories cannot achieve its maximal value of $\mathcal{C}=1$ and it exhibits a clear decay after the first jump event. This is because that the vacuum damping (the third terms in Eq.(\ref{drhop})), which models the inefficient detection, contaminates the pure Bell states. Hence, the longer waiting time for the first  ``click", the less amount of entanglement is obtained. Further, the entanglement no longer increases after the first ``click", and it then decays continuously whether or not photons are detected later, because subsequent ``clicks" do not alter the entanglement degree of the changing entangled states but the existing vacuum damping decreases the entanglement all the time. It should be noted that recent advances in superconducting nanowire single-photon detectors (SNSPDs) have already resulted in a detection efficiency close to 100$\%$ \cite{det}.

\subsection{Bell states via homodyne detection}
\begin{figure*}[t]
	\centering
	\includegraphics[bb=-100 50 1000 750,scale=0.6]{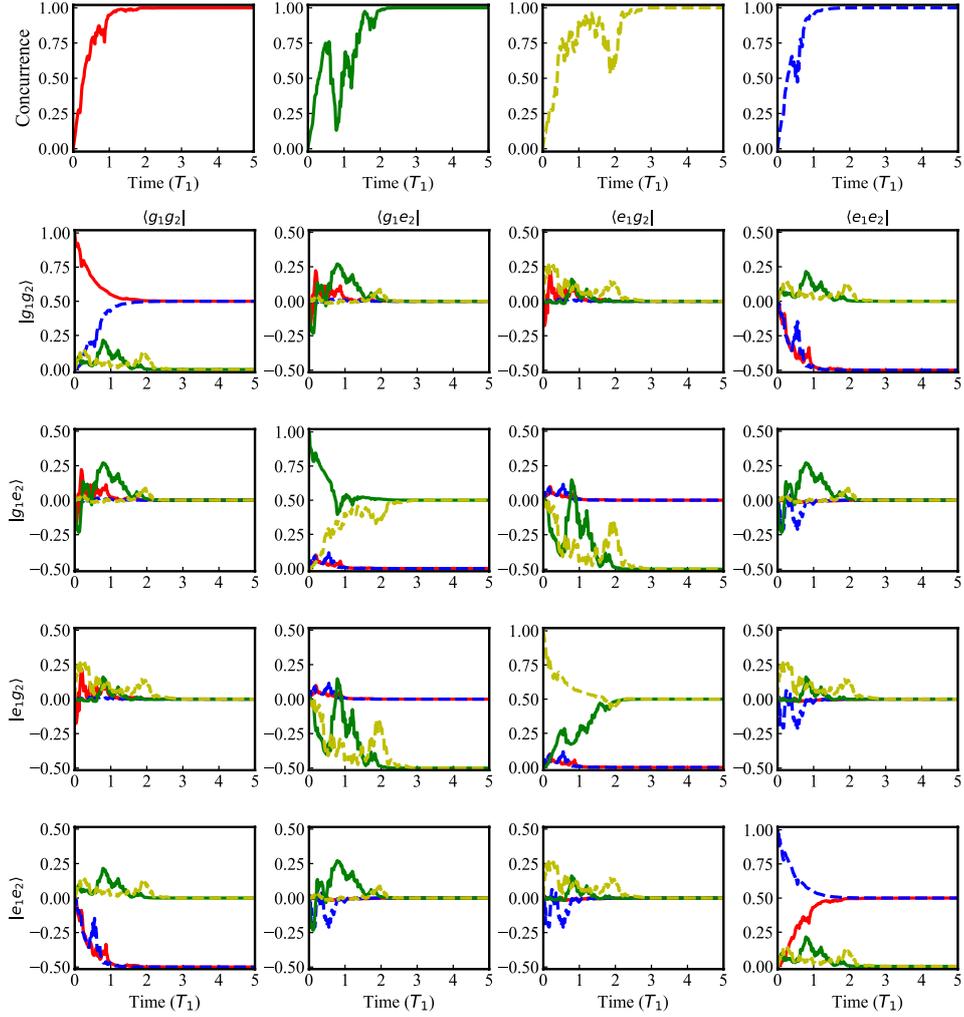}
    \caption{(the first row) Selected diffusive trajectories of the concurrence which reach the maximal values in the steady-state regime, for the initial states $|g_1g_2\rangle$ (red line), $|g_1e_2\rangle$ (green line), $|e_1g_2\rangle$ (yellow dashed line), and $|e_1e_2\rangle$ (blue dashed line). The plots in the second, third and fourth rows are the evolution of the elements of the corresponding density matrix of each trajectory. The parameters are same as Fig. {\ref 5}. }
	\label{6}
\end{figure*}
%\begin{figure*}[t]
	%\centering
	%\includegraphics[bb=50 80 800 520,scale=0.5]{0708fig7a.pdf}
	%\includegraphics[bb=50 80 800 530,scale=0.5]{0708fig7b.pdf}
	%\caption{(the first row) Selected diffusive trajectories of the concurrence which reach the maximal values in the steady-state regime, for the initial states $|g_1g_2\rangle$ (red line), $|g_1e_2\rangle$ (green line), $|e_1g_2\rangle$ (yellow dashed line), and $|e_1e_2\rangle$ (blue dashed line). The plots in the second, third and fourth rows are the evolution of the elements of the corresponding density matrix of each trajectory. The parameters are same as Fig. {\ref 5}. }
%	\label{6}
%\end{figure*}

In this subsection, we investigate the entanglement between the two qubits which are subject to continuous homodyne detection. We only consider one of the waveguide outputs (e.g. $\eta_l=0$) is monitored since the same results are reached when both are homodyned. Fig. \ref{5} depicts the concurrence of some selected diffusive trajectories and the average concurrence $\overline{\mathcal C}$ (dashed lines) over an ensemble of $2000$ quantum trajectories for different homodyne-detection efficiencies $\eta_r$, with the initial state $|g_1g_2\rangle$ and the distance $kd=2n\pi$. It is clearly shown that all trajectories demonstrate short-time entanglement, some trajectories display null long-time entanglement, while the other trajectories possess steady maximal entanglement $\mathcal{C}=1$ independent of the detection efficiency $\eta_r$. This can be understood as follows: it can be found from Eq.(\ref{rho}) the state $\tilde{\rho}_{\rm ss}=\frac{1}{2}|\Psi_-\rangle\langle \Psi_-|+\frac{1}{4}(|\phi_+\rangle\langle \phi_+|+|\phi_{-}\rangle\langle \phi_{-}|)$ is its possible solutions in the long-time regime, where the states
$|\phi_\pm\rangle=\frac{1}{2}\big[|e_1e_2\rangle+|g_1g_2\rangle\pm(| g_1 e_2\rangle+|e_1g_2\rangle
)\big]$. Because the states $|\Psi_-\rangle$ and $|\phi_\pm\rangle$ are the eigenstates of the operator $\hat J_{r+}$, satisfying $\hat J_{r+} |\Psi_-\rangle=0$ and $\hat J_{r+} |\phi_{\pm}\rangle=\pm |\phi_{\pm}\rangle$ and the unconditional equation (\ref{rho}) can be unraveled into a set of quantum trajectories governed by the conditional equation (\ref{drhoc}), the state $\tilde {\rho}_{\rm ss}$ can be considered as an ensemble average of the conditioned states $|\Psi_-\rangle$ and $|\phi_\pm\rangle$
of the quantum trajectories of the homodyne detection (involving in the measurement operator $\hat J_{r+}$) in the steady-state regime. Since the operator $\hat J_{r+}$ is hermitian, its eigenstates $|\Psi_-\rangle$ and $|\phi_\pm\rangle$ satisfy the conditional equation in the steady-state regime, even the eigenvalues are nonzero for the latter and the detection is inefficient ($\eta_r<1$). As a result, as shown in Fig.\ref{5}, the entanglement of the trajectories has the steady-state values of $\mathcal C=1$ or $\mathcal C=0$, which correspond to the states $|\Psi_-\rangle$ and $|\phi_\pm\rangle$, respectively. The steady Bell state $|\Psi_-\rangle$ can therefore be achieved with a fifty-percent probability, which is moreover immune to the detection inefficiency. This is distinct from the case of photon counting. For finite detection efficiency, the transient states of the trajectories are still mixed and the time for approaching the steady states is prolonged as $\eta_r$ decreases, since less output information about the spin of the two qubits has been accessed.

From the above discussion, the average entanglement $\overline {\mathcal C}=0.5$ should be achievable in the long-time regime, as shown in Fig.\ref{5}. The explicit expression for the average concurrence in the whole time can be derived from the stochastic equation (\ref{drhoc}) for $\eta_r=1$. For a pure two-qubit state $|\widetilde{\varphi}_c\rangle=\varphi_0|g_1g_2\rangle+\varphi_1|g_1e_2\rangle+\varphi_2|e_1g_2\rangle+\varphi_3|e_1e_2\rangle$., the concurrence is $\mathcal{C}=2|\varphi_0\varphi_3-\varphi_1\varphi_2|$, and thus the evolution of the concurrence $\mathcal C$ is derived as
\begin{align}
d\mathcal{C}(\widetilde{\varphi}_c)&=\big|-3\gamma\big [ \mathcal{C}(\widetilde{\varphi}_c)-(\varphi_1^2+\varphi_2^2-\varphi_0^2-\varphi_3^2)\big]dt\nonumber\\
&~~~-2\sqrt\gamma\mathcal{C}(\widetilde{\varphi}_c)\langle\hat J_{r+}\rangle dW_r\big|,
    %d\mathcal{C}(\widetilde{\varphi}_c)&=\big|-3\gamma\mathcal{C}(\widetilde{\varphi}_c)dt+3\gamma(\varphi_1^2+\varphi_2^2-\varphi_0^2-\varphi_3^2)dt\nonumber\\
	%&~~~-2\sqrt\gamma\mathcal{C}(\widetilde{\varphi}_c)\langle\hat J_{r+}\rangle dW_r\big|,
	\label{dc}
\end{align}
from which the ensemble average of the concurrence can be calculated as

\begin{align}
	\overline{\mathcal{C}}(\widetilde{\varphi}_c)=\frac{1}{2}-\frac{1}{5}e^{-3\gamma t}-\frac{3}{10}e^{-8\gamma t},
	\label{Ec}
\end{align}
for the initial state $|g_1g_2\rangle$, which coincides with the numerical result.

In Fig.\ref{6}, the entanglement of trajectories for different initial states are plotted. We specifically choose four individual trajectories whose concurrence $\mathcal C=1$ in the long-time limit, for the initial states $|g_1g_2\rangle$, $|g_1e_2\rangle$, $|e_1g_2\rangle$, and $|e_1e_2\rangle$, respectively. In addition, the corresponding populations and coherences are also plotted.  Note that similar to Fig.\ref{5}, the entanglement of the trajectories for these initial states also becomes $\mathcal C=1$ or $\mathcal C=0$ in the steady-state regime. It is shown that the individual trajectory initialized at $|g_1g_2\rangle$ (red line) or $|e_1e_2\rangle$ (blue-dashed line) eventually evolves into the Bell state $|\Psi_-\rangle$, while the trajectory started from  $|g_1e_2\rangle$ (green line) or $|e_1g_2\rangle$ (yellow-dashed line) asymptotically approaches the another Bell state $|\Phi_-\rangle$. This is different from the cyclic jumps in the former case of photon counting. It should be pointed out that if the distance of the two qubits satisfies $kd=(2n+1)\pi$, the steady Bell states $|\Psi_+\rangle$ or $|\Phi_+\rangle$ of the trajectories can be resulted, respectively for the initial states $|g_1g_2\rangle$ ($|e_1e_2\rangle$) or $|e_1g_2\rangle$ ($|g_1e_2\rangle$). Therefore, from the discussion, continuous homodyne measurement can also allow us to probabilistically generate long-time Bell states via making a post-selection on the diffusive trajectories with the long-time concurrence $\mathcal{C}=1$.

\section{Conclusion}
In conclusion, we show in this paper how to prepare long-term sustainable Bell states of two distant qubits by using time-continuous photon counting or homodyne detection. We consider two identical $\Lambda$ emitters which are coupled to a one-dimension waveguide via off-resonant Raman scattering. It is shown that in both of detection schemes, Bell states can be realized in the long-time regime. For the case of photon counting,  a cyclic jump among Bell states can be formed and the alternate appearance of different Bell states is heralded on the subsequent photon-detection events in the long-time regime. While for the case of homodyne detection, it is found that different Bell states can be achieved in the regime of steady states with a probability of fifty percent, independent of detection efficiency. Our scheme is advantageous over previous ones in which transient or intermittent Bell states of qubits can only be generated, and it may find applications in e.g. quantum communication networks.

\section*{Acknowledgment}
This work is supported by the National Natural Science Foundation of China (No.11674120) and the Fundamental Research Funds for the Central Universities (No. CCNU18TS033).

\bibliographystyle{plain}

\end{document}